\documentclass[twocolumn,amsmath,amssymb,aps,prb,floatfix,showpacs,citeautoscript]{revtex4-1}

\usepackage{graphicx}
\usepackage{bm}
\usepackage{bbold} %

\usepackage[utf8]{inputenc}
\usepackage[T1]{fontenc}

\newcommand{\equ}[1]{Eq.~(\ref{eq:#1})}

\def\ket#1{\vert#1\rangle}
\def\bra#1{\langle#1\vert}

\def\beq{\begin{equation}}
\def\eeq{\end{equation}}

\def\P{{\bf P}}
\def\M{{\bf M}}
\def\eb{\bm{\mathcal{E}}}
\def\ec{\mathcal{E}}
\def\bb{{\bf B}}

\def\z2{$\mathbb{Z}_2$}

\def\inv{{\cal P}}
\def\timer{{\cal T}}

\def\k{{\bf k}}

\def\im{\mathrm{Im}}

\def\tr{\mathrm{tr}}

\def\i{{\bf i}}

\def\k{{\bf k}}

\def\hk{H^0_\k}
\def\h0{H^0}

\def\wt#1{\widetilde{#1}}

\def\part{\wt{\partial}}

\begin{document}

\title{Orbital magnetoelectric coupling at finite electric field}

\author{Andrei Malashevich}
\email{andreim@berkeley.edu}
\affiliation{
Department of Physics, University of California, Berkeley, California 94720, USA
}
\author{David Vanderbilt}
\affiliation{
Department of Physics \& Astronomy, 
Rutgers, the State University of New Jersey, 
Piscataway, New Jersey 08854, USA
}
\author{Ivo Souza}
\affiliation{
Centro de F\'{\i}sica de Materiales and DIPC, Universidad del Pa\'\i s Vasco, 
20018 San Sebasti\'an, Spain
}
\affiliation{Ikerbasque, Basque Foundation for Science, E-48011 Bilbao, Spain}

\date{\today}

\begin{abstract}
  We extend the band theory of linear orbital magnetoelectric coupling
  to treat crystals under finite electric fields. Previous work
  established that the orbital magnetoelectric response of a
  generic insulator at zero field comprises three contributions that
  were denoted as local circulation, itinerant circulation, and
  Chern-Simons. We find that the expression for each of them is
  modified by the presence of a dc electric field. Remarkably, the sum
  of the three correction terms vanishes, so that the total coupling
  is still given by the same formula as at zero field.  This
  conclusion is confirmed by numerical tests on a tight-binding model,
  for which we calculate the field-induced change in the linear
  magnetoelectric coefficient.
\end{abstract}

\pacs{75.85.+t,03.65.Vf,71.20.Ps}

\maketitle
\marginparwidth 2.7in
\marginparsep 0.5in
\def\scr{\scriptsize}

Magnetoelectrics are magnetic insulators whose dielectric polarization
$\P$ changes linearly under a small applied magnetic field $\bb$ and,
conversely, whose magnetization $\M$ changes linearly with a small
applied electric field $\eb$.\cite{odell70,fiebig05} This linear
magnetoelectric (ME) coupling is described by the response
tensor\cite{explan-units}
\beq
\label{eq:me}
\alpha_{ij}=\frac{\partial M_j}{\partial\ec_i}
=\frac{\partial P_i}{\partial B_j},
\eeq
which is odd under both spatial inversion ($\inv$) and
time-reversal ($\timer$) symmetries. Thus ME materials must be acentric and
display magnetic order.

In crystals where only one of the two symmetries, $\inv$ or $\timer$, is
present, it may still be possible to induce a linear ME effect
by applying an external field which breaks that symmetry.
So, for example, a centrosymmetric
insulating antiferromagnet placed in a (strong) electric field loses
its inversion center. Likewise, a nonmagnetic ferroelectric crystal 
loses time-reversal symmetry when subject to a magnetic field.
In both cases the symmetry is sufficiently lowered that the
tensor $\alpha$ becomes nonzero.

It is useful to view these field-induced effects as higher-order ME
responses of the unbiased crystal.~\cite{ascher-philmag68} Two
quad\-rat\-ic ME effects can be defined in this way. Going to
next order in magnetic field yields the tensor
\beq
\label{eq:beta}
\beta_{ijk}=\frac{\partial\alpha_{ij}}{\partial B_k}
=\frac{\partial^2 P_i}{\partial B_j\partial B_k},
\eeq
which is odd under $\inv$ and even under $\timer$.
Going instead to next order in the electric field gives
\beq
\label{eq:gamma}
\gamma_{ijk}=\frac{\partial\alpha_{ji}}{\partial \ec_k}
=\frac{\partial^2 M_i}{\partial\ec_j\partial\ec_k},
\eeq
which is even under $\inv$ and odd under $\timer$.
Reference~\onlinecite{ascher-philmag68} lists the form of these tensors
for all the crystal classes. 
While most 
investigations of ME couplings in solids have focused on the linear
response $\alpha$ for a reference state of the crystal at 
zero electric and magnetic fields,
the quadratic responses $\beta$ and $\gamma$ have also been measured
in materials where $\alpha$ vanishes by symmetry. In particular the
electric-field-induced effect, which constitutes the primary focus of
this work, was first measured by O'Dell in yttrium
iron garnet.\cite{odell-philmag67}

The ME response can be divided into
four contributions, depending on whether the response is
frozen-ion (purely electronic) or lattice-mediated,
and whether it is spin
or orbital in character. We will refer to the frozen-ion
part of the orbital response as the orbital magnetoelectric polarizability 
(OMP).\cite{essin09,malash2010}
While the OMP is typically a small contribution to the ME
response in conventional magnetoelectrics, it was recently realized
that, under certain conditions of surface preparation, \z2-odd
topological insulators\cite{hasan-rmp10} should display a large,
quantized OMP response.\cite{qi08,essin09} This is a remarkable
prediction, especially considering that in this class of materials
$\timer$ symmetry is preserved in the bulk (it must, however, be
broken on the surface). This topological magnetoelectric
effect has triggered a great deal of interest in orbital
magnetoelectric couplings in solids.

The microscopic theory needed to calculate the OMP 
at zero electric and magnetic fields
from first principles was worked out in
Refs.~\onlinecite{malash2010,essin2010}.  In addition to the so-called
Chern-Simons term responsible for the topological ME
effect,\cite{qi08,essin09,coh}
it was found that two more (Kubo) terms
contribute to the OMP in conventional magnetoelectrics in which $\timer$
and $\inv$ symmetries are broken spontaneously in the bulk.

In this work we generalize the band theory 
of OMP of periodic insulators\cite{malash2010,essin2010} 
to finite electric fields. That is, we evaluate the coefficient
$\alpha$ at nonzero $\eb$,
\beq
\label{eq:me-e}
\alpha_{ij}(\eb)=\left.\frac{\partial M_j}{\partial \ec_i}\right|_{\bb=0}.
\eeq
(Henceforth, the condition $\bb=0$ will be implied throughout. 
It is also understood that from now on $\alpha$ denotes
the OMP part of the entire ME response.) 
A principal result of our work is the conclusion 
that the zero-field expression for the {\it total} OMP remains valid
at finite electric field, while the above-mentioned Chern-Simons and Kubo
terms separately acquire field-induced contributions.
We confirm our formal results by numerical tests on a tight-binding model.

Our derivation of a formula for $\alpha(\eb)$
proceeds along the lines of Ref.~\onlinecite{malash2010}.  We start
from the expression given therein for the orbital magnetization
of a generic band insulator under a finite electrical bias. 
It comprises three terms,
\beq
\label{eq:m-e}
M_j(\eb)=M^{\rm LC}_j(\eb)+M^{\rm IC}_j(\eb)+M^{\rm CS}_j(\eb),
\eeq
where
\beq
\label{eq:m-lc}
M^{\rm LC}_j=-\frac{\eta}{2}\epsilon_{jpq}
\int d^3k\,\im\,\bra{\wt{\partial}_pu_{n\k}}\hk
\ket{\wt{\partial}_qu_{n\k}},
\eeq
\beq
\label{eq:m-ic}
M^{\rm IC}_j=-\frac{\eta}{2}\epsilon_{jpq}
\int d^3k\,\im\,\left\{\bra{u_{n\k}}\hk\ket{u_{m\k}}
\bra{\wt{\partial}_pu_{m\k}}
\wt{\partial}_qu_{n\k}\rangle\right\},
\eeq
and
\beq
\label{eq:m-cs}
M^{\rm CS}_j=\frac{e\eta}{2}\ec_j
\int d^3k\,
\epsilon_{pqr}
\tr\left[A_p\partial_qA_r-\frac{2i}{3}A_pA_qA_r\right].
\eeq
The common prefactor in these formulas is $\eta=-e/\hbar(2\pi)^3$
($e>0$ is the magnitude of the electron charge), and a sum is
implied over repeated Cartesian ($pqr$) and valence-band
($mn$) indices. The cell-periodic part of the field-polarized Bloch
state\cite{souza02} is denoted by $\ket{u_{n\k}}$, 
$\partial_j$ is the partial derivative
with respect to the $j$th component of the wavevector $\k$,
and the tilde
indicates a covariant derivative $\wt{\partial}_j= Q_\k\partial_j$,
where $Q_\k=1-\ket{u_{n\k}}\bra{u_{n\k}}$ (sum implied over $n$).
The Hamiltonian $\hk$ is defined as
\beq
\hk=e^{-i\k\cdot\bf r}{\cal H}^0e^{i\k\cdot\bf r},
\eeq
where ${\cal H}^0$ is the zero-field part of the crystal Hamiltonian.
In \equ{m-cs} the symbol $A_p$ denotes the Berry connection matrix
\beq
A_{mn\k p}=i\bra{u_{m\k}}\partial_p u_{n\k}\rangle, 
\eeq
and the trace is over the valence bands. 

Equations (\ref{eq:m-lc}) and (\ref{eq:m-ic}) describe respectively
the {\it local} and {\it itinerant} circulation contributions to
the magnetiztion,\cite{malash2010} while \equ{m-cs} is the
Chern-Simons term. At variance with the other two terms, whose
dependence on the electric field is only implicit, ${\bf M}^{\rm
CS}$ displays an explicit linear dependence on $\eb$. It is
therefore expedient to introduce a new quantity $M^{\rm CS}_1$ via
the relation
\beq
\label{eq:m-cs-e}
M^{\rm CS}_j(\eb)\equiv\ec_jM^{\rm CS}_1(\eb),
\eeq
where the subscript `1' serves as a reminder that $M^{\rm CS}_1$
enters the expression for ${\bf M}$ multiplied by $\eb$ to the
first power.

All three magnetization terms, 
${\bf M}^{\rm LC}$, ${\bf M}^{\rm IC}$, and ${\bf M}^{\rm CS}$,
are invariant under gauge
transformations within the valence-band manifold, although in the case
of ${\bf M}^{\rm CS}$ this invariance is only modulo a quantum of
indeterminacy.\cite{qi08} In the limit that $\eb$ goes to zero, 
${\bf M}^{\rm CS}$ vanishes and \equ{m-e} reduces to the expression for
the spontaneous orbital magnetization.\cite{ceresoli06}

As already mentioned, all terms in \equ{m-e} can contribute to
the linear ME coupling, Eq.~(\ref{eq:me-e}), so that
\beq
\label{eq:me-tot}
\alpha_{ij}(\eb)=\alpha^{\rm LC}_{ij}(\eb)+
                 \alpha^{\rm IC}_{ij}(\eb)+
                 \alpha^{\rm CS}_{ij}(\eb).
\eeq
The derivation of the expressions for these objects is straightforward
though somewhat lengthy. It essentially repeats the steps in
Appendix~B of Ref.~\onlinecite{malash2010}, where the derivation was
carried out for the LC and IC (``Kubo'') terms under the assumption
that $\eb=0$ (the CS term is trivial at $\eb=0$).  At $\eb\not=0$ one
may show that each of the terms in \equ{me-tot} consists of a
``zero-field'' part plus a ``field-correction'' part having an
explicit linear dependence on $\eb$,
\beq
\label{eq:fld-corr}
\alpha_{ij}(\eb)=\alpha_{0,ij}(\eb)+\ec_j\alpha_{1,i}(\eb).
\eeq
The field-correction terms for the LC and IC contributions can be traced back to
Eqs.~(B.7) and (B.8) in Ref.~\onlinecite{malash2010}, 
which at $\eb\not=0$ acquire extra terms. 
As for the Chern-Simons contribution, differentiating \equ{m-cs-e}
with respect to $\ec_j$ yields $\alpha^{\rm CS}_{0,ij}=\delta_{ij}M_1^{\rm CS}$
and $\alpha_{1,i}^{\rm CS}=\partial M_1^{\rm CS}/\partial\ec_i$.

Thus, we arrive at the results
\beq
\label{eq:omp0-lc}
\begin{split}
  \alpha^{\mathrm{LC}}_{0,ij}(\eb)=
  \eta\epsilon_{jpq}\mathrm{Im}\int d^3k\, \Big(
  &\bra{\wt{\partial}_p u_{n\k}}(\partial_q\hk)\ket{\wt{D}_i u_{n\k}}\\
  -&\frac{1}{2}\bra{\wt{\partial}_p u_{n\k}}(D_i\hk)
  \ket{\wt{\partial}_qu_{n\k}} \Big),
\end{split}
\eeq
\beq
\label{eq:omp0-ic}
\begin{split}
\alpha^{\mathrm{IC}}_{0,ij}(\eb)=
\eta\epsilon_{jpq}\mathrm{Im}\int &d^3k\,
  \Big(
      \bra{\wt{\partial}_p u_{n\k}}\wt{D}_i u_{m\k}\rangle
\bra{u_{m\k}}(\partial_q\hk)\ket{u_{n\k}}\\
     -&\frac{1}{2}\bra{\wt{\partial}_p u_{n\k}}\wt{\partial}_q u_{m\k}\rangle
       \bra{u_{m\k}}(D_i \hk)\ket{u_{n\k}}
\Big),
\end{split}
\eeq
\beq
\label{eq:omp0-cs}
\alpha^{\mathrm{CS}}_{0,ij}(\eb)=\delta_{ij}
\eta\frac{e}{2}\int d^3k\,\epsilon_{pqr}
\tr\left[A_p\partial_qA_r-\frac{2i}{3}A_pA_qA_r\right],
\eeq
\beq
\label{eq:delta-lc}
\alpha^{\rm LC}_{1,i}(\eb)=\eta e\int d^3k\,\epsilon_{pqr}{\rm Re}
\left[
  \bra{\wt{D}_i u_{n\k}}\wt{\partial}_pu_{m\k}\rangle
  \bra{\wt{\partial}_q u_{m\k}}\wt{\partial}_r u_{n\k}\rangle
\right], 
\eeq
and
\beq
\label{eq:deltas}
\alpha^{\rm LC}_{1,i}(\eb)=\alpha^{\rm IC}_{1,i}(\eb)=-\frac12\alpha^{\rm CS}_{1,i}.
\eeq
In the above expressions, $D_i$ is the partial derivative with respect
to the $i$th component of the electric field.  The terms containing
$D_i\hk$ in Eqs.~(\ref{eq:omp0-lc}) and (\ref{eq:omp0-ic}) are
screening corrections which are present in self-consistent
calculations.  

Equations (\ref{eq:omp0-lc})--(\ref{eq:omp0-cs}) for the
zero-field terms are essentially rewritten from
Ref.~\onlinecite{malash2010}. It should be emphasized, however, that
in the present context these expressions depend on the electric field
implicitly via the wave functions. The explicit field dependence is
given by the field-correction terms, Eqs.~(\ref{eq:delta-lc}) and
(\ref{eq:deltas}).  Remarkably, these terms are not independent and
add up to zero when inserted into \equ{me-tot}.  We conclude,
therefore, that the expression for the {\it total} OMP derived in
Refs.~\onlinecite{malash2010,essin2010} assuming $\eb=0$ remains valid
for $\eb\not=0$.  This constitutes one of our principal results.  The
explicit expression given in \equ{delta-lc} for the field-correction
terms is the other main result of this work.  It is useful if one is
interested in the field dependence of the separate gauge-invariant
contributions to the OMP.  Because it contains three $k$ derivatives
and one field derivative, this quantity is even under $\inv$ and odd
under $\timer$, just like the coefficient $\gamma$ defined by
\equ{gamma}. This is reasonable since, as one can see from
\equ{fld-corr}, $\alpha_1^{\rm LC/IC/CS}$ gives a contribution to
$\gamma^{\rm {LC/IC/CS}}$ and should therefore have the same symmetry
properties.

As a check of our analytic derivation, we have implemented the formula
for $\alpha(\eb)$ in a tight-binding model, and used it to calculate
the nonlinear ME coefficient
$\gamma_{zzz}$ at $\eb=0$. Since the tensor $\gamma$
vanishes in $\timer$-invariant systems, we need a model where
$\timer$ is spontaneously broken, and we chose that of
Ref.~\onlinecite{malash2010}.
This is a spinless model with eight sites per primitive cell
arranged on a $2\times2\times2$ cube, where $\timer$ symmetry is
broken by complex nearest-neighbor hoppings, and we have used the same
on-site energies and nearest-neighbor hoppings tabulated in that
work. (This choice of parameters also breaks $\inv$, so that the
linear ME tensor $\alpha$ is nonzero already at $\eb=0$, but this is
not essential for our present purposes.)  As in
Ref.~\onlinecite{malash2010} the two lowest bands were treated as
occupied, and the phase $\varphi$ of one of the complex hoppings was
chosen as a control parameter for plotting purposes.

The technical details of the tight-binding implementation of
Eqs.~(\ref{eq:m-lc})--(\ref{eq:m-cs}) and (\ref{eq:omp0-lc})--(\ref{eq:delta-lc})
can be found in Ref.~\onlinecite{malash2010}.  The only significant
difference with respect to that work is that the field derivative
$\ket{\wt{D}_i u_{n\k}}$ of the cell-periodic Bloch states must be
evaluated at {\it finite} $\eb$. Under these circumstances the usual
``sum-over-states'' formula~\cite{ceresoli06} cannot be employed,
and one must instead minimize a suitably defined
functional.~\cite{wang-prb07}

We shall calculate the $zzz$ component of $\gamma$ 
from the first equality in \equ{gamma}.
Combining with \equ{me-tot} we find
\beq 
\label{eq:gamma-lciccs}
\gamma=\gamma^{\rm LC}+\gamma^{\rm IC}+\gamma^{\rm CS}.  
\eeq 
The CS term is the simplest to evaluate, as the derivative of
\equ{fld-corr} with respect to $\ec_z$ can be taken
analytically. 
The zero-field and field-correction terms therein both contribute an
amount $\alpha^{\rm CS}_{1,z}(0)$ to $\gamma_{zzz}^{\rm CS}(0)$.
Thus,
\beq
\label{eq:zzz-cs}
\gamma_{zzz}^{\rm CS}(0)=
2\alpha^{\rm CS}_{1,z}(0)=-4\alpha^{\rm LC}_{1,z}(0),
\eeq
where the second equality follows from
\equ{deltas}. The quantity on the right-hand side can be evaluated
directly from \equ{delta-lc}.  For the LC and IC terms we calculate
the derivative of the zero-field terms in \equ{fld-corr} using finite
differences and obtain
\beq
\label{eq:zzz-1}
\gamma_{zzz}^{\rm LC/IC}(0)
\simeq \frac{\alpha_{0,zz}^{\rm LC/IC}(\ec_z)-\alpha_{0,zz}^{\rm
    LC/IC}(-\ec_z)}{2\ec_z} +\alpha^{\rm LC}_{1,z}(0).  
\eeq
In practice we evaluate the first term from Eqs.~(\ref{eq:omp0-lc})
and (\ref{eq:omp0-ic}), using small positive and negative fields along
$z$ of magnitude $\ec_z=1.0\times10^{-5}\,\mathrm{V}/\mathrm{m}$.

The results of the above calculations were compared with a 
finite-difference determination of the second field derivative of 
${\bf M}$,
\beq
\label{eq:zzz-2}
\gamma_{zzz}(0)=
\left.\frac{\partial^2 M_z}{\partial \ec_z^2}\right|_{\eb=0}
\simeq
\frac{M_z(\ec_z)-2M_z(0)+M_z(-\ec_z)}{\ec_z^2},
\eeq
using the $k$-space expressions from
Ref.~\onlinecite{malash2010} for the LC, IC and CS terms in \equ{m-e}.
The results obtained in this manner can be taken as a reference,
since the $k$-space expression for ${\bf M}(\eb)$ has been carefully
tested by comparing with real-space calculations on bounded samples
cut from the bulk crystal.\cite{malash2010}

\begin{figure}
\centering\includegraphics{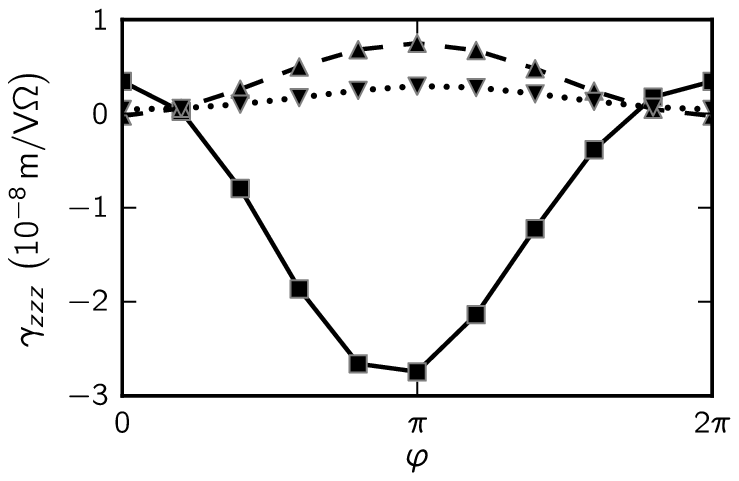}
\caption{Decomposition of $\gamma_{zzz}$ of \equ{gamma-lciccs} 
into $\gamma^{\mathrm{LC}}$ (solid lines), $\gamma^{\mathrm{IC}}$
(dashed lines), and $\gamma^{\mathrm{CS}}$ (dotted lines) calculated
using Eqs.~(\ref{eq:zzz-cs}) and (\ref{eq:zzz-1}). Symbols denote the 
same contributions evaluated using \equ{zzz-2}.
}
\label{fig:gamma_zzz}
\end{figure}

The agreement between the two
sets of calculations can be seen in Fig.~\ref{fig:gamma_zzz}, where
the LC, IC, and CS contributions to $\gamma_{zzz}$ are plotted
separately as functions of $\varphi$. In this calculation
$\gamma^{\rm CS}_{zzz}$ is about an order of magnitude smaller than
$\gamma^{\rm LC}_{zzz}$.  From Eqs.~(\ref{eq:zzz-cs}) and (\ref{eq:zzz-1})
it then follows that the field-correction terms contribute little,
especially in the case of $\gamma^{\rm LC}_{zzz}$.  Further numerical
tests focusing on those small terms are therefore desirable.

\begin{figure}
\centering\includegraphics{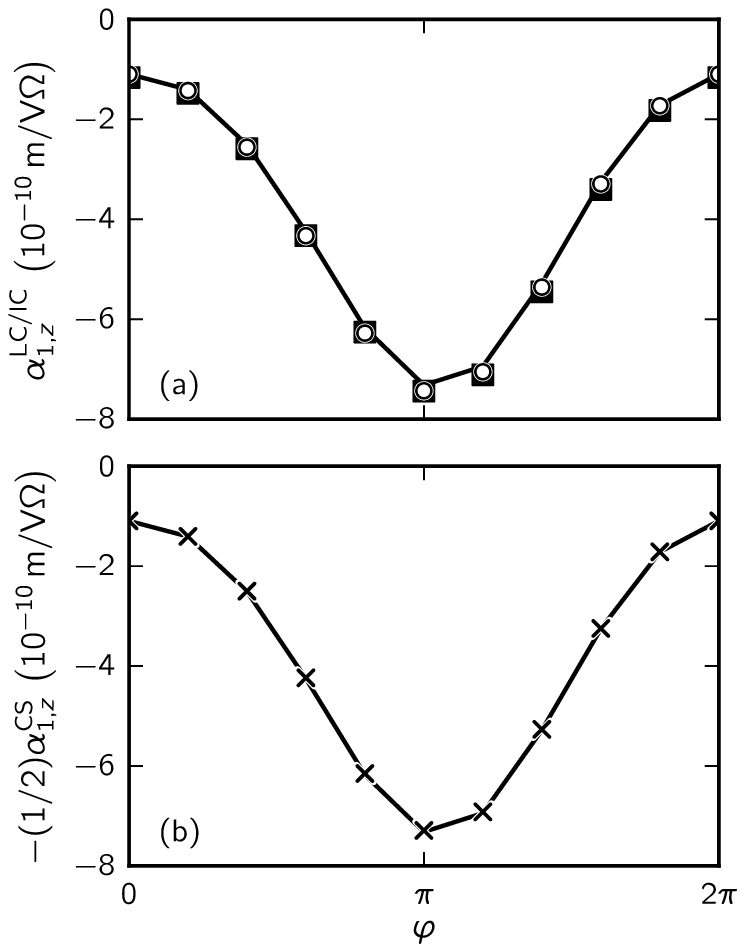}
\caption{(a) Right-hand side (solid line) and left-hand side (symbols)
  of \equ{visual-check}. Squares and circles denote the LC and IC
  contributions, respectively.  (b) Equation~(\ref{eq:zzz-cs}) (solid
  line) and \equ{zzz-2} for the CS contribution (crosses),
  both multiplied by a factor of $-1/4$ for visual check of
  \equ{deltas} by comparison to (a).}
\label{fig:beta_zzz}
\end{figure}

In order to isolate the field-correction terms in 
$\gamma^{\rm LC}_{zzz}$ and $\gamma^{\rm IC}_{zzz}$, 
we subtract the zero-field terms from the total:
\beq
\label{eq:visual-check}
\left[
  \frac{\partial^2 M_z^{\rm LC/IC}}{\partial\ec_z^2}-
  \frac{\partial\alpha_{0,zz}^{\rm LC/IC}}{\partial\ec_z}
\right]_{\eb=0}=
\alpha^{\rm LC/IC}_{1,z}(0).
\eeq
In Fig.~\ref{fig:beta_zzz}~(a) we plot, as a function
of $\varphi$, the two sides of this equation. The
field derivatives on the left-hand side are evaluated by finite
differences, while the right-hand side is calculated from
\equ{delta-lc}.  It is clear that the field-correction terms in
\equ{fld-corr} are nonzero, and the good agreement between the three
curves demonstrates that for both LC and IC they are given by
\equ{delta-lc}.

The CS contribution does
not need additional tests since, as noted above, the
contributions to $\gamma^{\rm CS}_{zzz}$ from the zero-field and
field-correction terms are identical.
However, we reproduce in Fig.~\ref{fig:beta_zzz}~(b)
the CS curve from Fig.~\ref{fig:gamma_zzz} multiplied by a factor
$-1/4$, so that the correctness of \equ{deltas} can be verified by
direct visual inspection.  This completes the numerical checks of
the $k$-space formula for $\alpha(\eb)$.

To summarize, we have extended the recently developed band theory of
orbital magnetoelectric response to treat crystals under a finite 
electrical bias. 
The theory presented in this work may be especially useful in calculations
of the second-order magnetoelectric effect defined by \equ{gamma}.
While it is possible in principle to calculate the second derivative
of ${\bf M}$ by finite differences, the numerical stability is likely
to be improved by taking one of the field derivatives
analytically, leaving only one derivative to be performed numerically.
We have demonstrated that in order to calculate the {\it total} OMP at
finite electric field, one may use the same 
equations (\ref{eq:omp0-lc})--(\ref{eq:omp0-cs}) that were
previously derived for zero field.  This is true even though the
individual local-circulation, itinerant-circulation, and Chern-Simons
contributions do separately acquire field-correction terms.  At
present, we are not aware of any simple argument that could have
anticipated the exact cancellation of these terms in the expression
for the total OMP.

We thank Sinisa Coh for stimulating discussions.
The work was supported by NSF 
under Grants No. DMR-0706493 and No. DMR-1005838.
Computational resources were provided by NERSC.


\begin{thebibliography}{10}%
\makeatletter
\providecommand \@ifxundefined [1]{%
 \ifx #1\undefined \expandafter \@firstoftwo
 \else \expandafter \@secondoftwo
\fi
}%
\providecommand \@ifnum [1]{%
 \ifnum #1\expandafter \@firstoftwo
 \else \expandafter \@secondoftwo
\fi
}%
\providecommand \enquote [1]{``#1''}%
\providecommand \bibnamefont  [1]{#1}%
\providecommand \bibfnamefont [1]{#1}%
\providecommand \citenamefont [1]{#1}%
\providecommand\href[0]{\@sanitize\@href}%
\providecommand\@href[1]{\endgroup\@@startlink{#1}\endgroup\@@href}%
\providecommand\@@href[1]{#1\@@endlink}%
\providecommand \@sanitize [0]{\begingroup\catcode`\&12\catcode`\#12\relax}%
\@ifxundefined \pdfoutput {\@firstoftwo}{%
 \@ifnum{\z@=\pdfoutput}{\@firstoftwo}{\@secondoftwo}%
}{%
 \providecommand\@@startlink[1]{\leavevmode}%
 \providecommand\@@endlink[0]{}%
}{%
 \providecommand\@@startlink[1]{%
  \leavevmode
  \pdfstartlink
   attr{/Border[0 0 1 ]/H/I/C[0 1 1]}%
   user{/Subtype/Link/A<</Type/Action/S/URI/URI(#1)>>}%
  \relax
 }%
 \providecommand\@@endlink[0]{\pdfendlink}%
}%
\providecommand \url  [0]{\begingroup\@sanitize \@url }%
\providecommand \@url [1]{\endgroup\@href {#1}{\urlprefix}}%
\providecommand \urlprefix [0]{URL }%
\providecommand \Eprint[0]{\href }%
\@ifxundefined \urlstyle {%
  \providecommand \doi [1]{doi:\discretionary{}{}{}#1}%
}{%
  \providecommand \doi [0]{doi:\discretionary{}{}{}\begingroup
  \urlstyle{rm}\Url }%
}%
\providecommand \doibase [0]{http://dx.doi.org/}%
\providecommand \Doi[1]{\href{\doibase#1}}%
\providecommand \bibAnnote [3]{%
  \BibitemShut{#1}%
  \begin{quotation}\noindent
    \textsc{Key:}\ #2\\\textsc{Annotation:}\ #3%
  \end{quotation}%
}%
\providecommand \bibAnnoteFile [2]{%
  \IfFileExists{#2}{\bibAnnote {#1} {#2} {\input{#2}}}{}%
}%
\providecommand \typeout [0]{\immediate \write \m@ne }%
\providecommand \selectlanguage [0]{\@gobble}%
\providecommand \bibinfo [0]{\@secondoftwo}%
\providecommand \bibfield [0]{\@secondoftwo}%
\providecommand \translation [1]{[#1]}%
\providecommand \BibitemOpen[0]{}%
\providecommand \bibitemStop [0]{}%
\providecommand \bibitemNoStop [0]{.\EOS\space}%
\providecommand \EOS [0]{\spacefactor3000\relax}%
\providecommand \BibitemShut [1]{\csname bibitem#1\endcsname}%
\bibitem{odell70}%
  \BibitemOpen
  \bibfield{author}{%
  \bibinfo {author} {\bibfnamefont{T.}~\bibnamefont{O'Dell}},\ }%
  \emph{\bibinfo {title} {The Electrodynamics of Magneto-Electric Media}}\
  (\bibinfo {publisher} {North-Holland},\ \bibinfo {address} {Amsterdam},\
  \bibinfo {year} {1970})%
  \bibAnnoteFile{NoStop}{odell70}%
\bibitem{fiebig05}%
  \BibitemOpen
  \bibfield{author}{%
  \bibinfo {author} {\bibfnamefont{M.}~\bibnamefont{Fiebig}},\ }%
  \bibfield{journal}{%
  \bibinfo {journal} {J. Phys. D}\ }%
  \textbf{\bibinfo {volume} {38}},\ \bibinfo {pages} {R123} (\bibinfo {year}
  {2005})%
  \bibAnnoteFile{NoStop}{fiebig05}%
\bibitem{explan-units}%
  \BibitemOpen
  \bibinfo {note} {In this work we use SI units. For a discussion of various
  conventions and choices of units used to define a magnetoelectric tensor, see
  Ref.~\onlinecite{coh} and references therein.}%
  \bibAnnoteFile{Stop}{explan-units}%
\bibitem{ascher-philmag68}%
  \BibitemOpen
  \bibfield{author}{%
  \bibinfo {author} {\bibfnamefont{E.}~\bibnamefont{Ascher}},\ }%
  \bibfield{journal}{%
  \bibinfo {journal} {Philos. Mag.}\ }%
  \textbf{\bibinfo {volume} {17}},\ \bibinfo {pages} {149} (\bibinfo {year}
  {1968})%
  \bibAnnoteFile{NoStop}{ascher-philmag68}%
\bibitem{odell-philmag67}%
  \BibitemOpen
  \bibfield{author}{%
  \bibinfo {author} {\bibfnamefont{T.~H.}\ \bibnamefont{O'Dell}},\ }%
  \bibfield{journal}{%
  \bibinfo {journal} {Philos. Mag.}\ }%
  \textbf{\bibinfo {volume} {16}},\ \bibinfo {pages} {487} (\bibinfo {year}
  {1967})%
  \bibAnnoteFile{NoStop}{odell-philmag67}%
\bibitem{essin09}%
  \BibitemOpen
  \bibfield{author}{%
  \bibinfo {author} {\bibfnamefont{A.~M.}\ \bibnamefont{Essin}}, \bibinfo
  {author} {\bibfnamefont{J.~E.}\ \bibnamefont{Moore}},\ and\ \bibinfo {author}
  {\bibfnamefont{D.}~\bibnamefont{Vanderbilt}},\ }%
  \bibfield{journal}{%
  \bibinfo {journal} {Phys. Rev. Lett.}\ }%
  \textbf{\bibinfo {volume} {102}},\ \bibinfo {pages} {146805} (\bibinfo {year}
  {2009})%
  \bibAnnoteFile{NoStop}{essin09}%
\bibitem{malash2010}%
  \BibitemOpen
  \bibfield{author}{%
  \bibinfo {author} {\bibfnamefont{A.}~\bibnamefont{Malashevich}}, \bibinfo
  {author} {\bibfnamefont{I.}~\bibnamefont{Souza}}, \bibinfo {author}
  {\bibfnamefont{S.}~\bibnamefont{Coh}},\ and\ \bibinfo {author}
  {\bibfnamefont{D.}~\bibnamefont{Vanderbilt}},\ }%
  \bibfield{journal}{%
  \bibinfo {journal} {New J. Phys.}\ }%
  \textbf{\bibinfo {volume} {12}},\ \bibinfo {pages} {053032} (\bibinfo {year}
  {2010})%
  \bibAnnoteFile{NoStop}{malash2010}%
\bibitem{hasan-rmp10}%
  \BibitemOpen
  \bibfield{author}{%
  \bibinfo {author} {\bibfnamefont{M.~Z.}\ \bibnamefont{Hasan}}\ and\ \bibinfo
  {author} {\bibfnamefont{C.~L.}\ \bibnamefont{Kane}},\ }%
  \bibfield{journal}{%
  \bibinfo {journal} {Rev. Mod. Phys.}\ }%
  \textbf{\bibinfo {volume} {82}},\ \bibinfo {pages} {3045} (\bibinfo {year}
  {2010})%
  \bibAnnoteFile{NoStop}{hasan-rmp10}%
\bibitem{qi08}%
  \BibitemOpen
  \bibfield{author}{%
  \bibinfo {author} {\bibfnamefont{X.-L.}\ \bibnamefont{Qi}}, \bibinfo {author}
  {\bibfnamefont{T.~L.}\ \bibnamefont{Hughes}},\ and\ \bibinfo {author}
  {\bibfnamefont{S.-C.}\ \bibnamefont{Zhang}},\ }%
  \bibfield{journal}{%
  \bibinfo {journal} {Phys. Rev. B}\ }%
  \textbf{\bibinfo {volume} {78}},\ \bibinfo {pages} {195424} (\bibinfo {year}
  {2008})%
  \bibAnnoteFile{NoStop}{qi08}%
\bibitem{essin2010}%
  \BibitemOpen
  \bibfield{author}{%
  \bibinfo {author} {\bibfnamefont{A.~M.}\ \bibnamefont{Essin}}, \bibinfo
  {author} {\bibfnamefont{A.~M.}\ \bibnamefont{Turner}}, \bibinfo {author}
  {\bibfnamefont{J.~E.}\ \bibnamefont{Moore}},\ and\ \bibinfo {author}
  {\bibfnamefont{D.}~\bibnamefont{Vanderbilt}},\ }%
  \bibfield{journal}{%
  \bibinfo {journal} {Phys. Rev. B}\ }%
  \textbf{\bibinfo {volume} {81}},\ \bibinfo {pages} {205104} (\bibinfo {year}
  {2010})%
  \bibAnnoteFile{NoStop}{essin2010}%
\bibitem{coh}%
  \BibitemOpen
  \bibfield{author}{%
  \bibinfo {author} {\bibfnamefont{S.}~\bibnamefont{Coh}}, \bibinfo {author}
  {\bibfnamefont{D.}~\bibnamefont{Vanderbilt}}, \bibinfo {author}
  {\bibfnamefont{A.}~\bibnamefont{Malashevich}},\ and\ \bibinfo {author}
  {\bibfnamefont{I.}~\bibnamefont{Souza}},\ }%
  \bibfield{journal}{%
  \bibinfo {journal} {Phys. Rev. B}\ }%
  \textbf{\bibinfo {volume} {83}},\ \bibinfo {pages} {085108} (\bibinfo {year}
  {2011})%
  \bibAnnoteFile{NoStop}{coh}%
\bibitem{souza02}%
  \BibitemOpen
  \bibfield{author}{%
  \bibinfo {author} {\bibfnamefont{I.}~\bibnamefont{Souza}}, \bibinfo {author}
  {\bibfnamefont{J.}~\bibnamefont{{\'I}{\~n}iguez}},\ and\ \bibinfo {author}
  {\bibfnamefont{D.}~\bibnamefont{Vanderbilt}},\ }%
  \bibfield{journal}{%
  \bibinfo {journal} {Phys. Rev. Lett.}\ }%
  \textbf{\bibinfo {volume} {89}},\ \bibinfo {pages} {117602} (\bibinfo {year}
  {2002})%
  \bibAnnoteFile{NoStop}{souza02}%
\bibitem{ceresoli06}%
  \BibitemOpen
  \bibfield{author}{%
  \bibinfo {author} {\bibfnamefont{D.}~\bibnamefont{Ceresoli}}, \bibinfo
  {author} {\bibfnamefont{T.}~\bibnamefont{Thonhauser}}, \bibinfo {author}
  {\bibfnamefont{D.}~\bibnamefont{Vanderbilt}},\ and\ \bibinfo {author}
  {\bibfnamefont{R.}~\bibnamefont{Resta}},\ }%
  \bibfield{journal}{%
  \bibinfo {journal} {Phys. Rev. B}\ }%
  \textbf{\bibinfo {volume} {74}},\ \bibinfo {pages} {024408} (\bibinfo {year}
  {2006})%
  \bibAnnoteFile{NoStop}{ceresoli06}%
\bibitem{wang-prb07}%
  \BibitemOpen
  \bibfield{author}{%
  \bibinfo {author} {\bibfnamefont{X.}~\bibnamefont{Wang}}\ and\ \bibinfo
  {author} {\bibfnamefont{D.}~\bibnamefont{Vanderbilt}},\ }%
  \bibfield{journal}{%
  \bibinfo {journal} {Phys. Rev. B}\ }%
  \textbf{\bibinfo {volume} {75}},\ \bibinfo {pages} {115116} (\bibinfo {year}
  {2007})%
  \bibAnnoteFile{NoStop}{wang-prb07}%
\end{thebibliography}
\end{document}